\documentclass[pre,preprint]{revtex4-1}

\usepackage{graphicx}
\graphicspath{{./figures/}}
\usepackage{amsmath,amssymb}
\usepackage{psfrag}

\newcommand{\vp}{\varphi}

\begin{document}

\title{Traveling chimeras in oscillator lattices with advective-diffusive coupling}

\author{L. Smirnov}
\affiliation{Department of Control Theory, Research and Education Mathematical Center ``Mathematics for Future Technologies'',
Nizhny Novgorod State University, Gagarin Av. 23, 603022, Nizhny Novgorod, Russia}
\affiliation{Institute of Applied Physics of the Russian Academy of Sciences, Ul’yanov Str. 46, 603950, Nizhny Novgorod, Russia}
\author{A. Pikovsky}
\affiliation{Institute of Physics and Astronomy, Potsdam University, 14476 Potsdam-Golm, Germany}

\begin{abstract}
We consider a one-dimensional array of phase oscillators coupled via an auxiliary complex field.
While in the seminal chimera studies by Kumamoto and Battogtokh only diffusion of the field was considered, we include advection which  makes the coupling left-right asymmetric.
Chimera starts to move and we demonstrate, that a weakly turbulent moving pattern appears.
It possesses a relatively large synchronous domain where the phases are nearly equal, and a more disordered domain where the local driving field is small.
For a dense system with a large number of oscillators, there are strong local correlations in the disordered domain, which at most places looks like a smooth phase profile.
We find also exact regular traveling wave chimera-like solutions of different complexity, but only some of them are stable.
\end{abstract}

\date{\today}
\maketitle

\section{Introduction}
Chimera patterns are intriguing structures combining order and disorder in oscillatory media~\cite{Panaggio-Abrams-15,Omelchenko-18,Omelchenko-Knobloch-19}.
Since their discovery by Kuramoto and Battogtokh~(KB) twenty years ago~\cite{Kuramoto-Battogtokh-02}, they attracted high attention both in applications and experimental realizations~\cite{Tinsley_etal-12,wickramasinghe2013spatially,Martens_etal-13,Totz_etal-18}, and in theoretical treatment~\cite{Abrams-Strogatz-04,Omelchenko_etal-08,Laing-09,Bordyugov-Pikovsky-Rosenblum-10,Omelchenko-13,xie2015chimera,Kemeth_etal-16,Smirnov-Osipov-Pikovsky-17}.
Chimeras manifest themselves in synchronous and asynchronous patches, which are best characterized in terms of the oscillator phases.
Correspondingly, the most simple models are formulated in terms of the phase dynamics equations, and we will follow this approach in this paper.

A characteristic feature of chimeras is that they appear when coupling between oscillators is described by integral terms.
This allows for non-smooth in space phase profiles, where neighboring oscillators can be uncorrelated or weakly correlated.
Such disordered profiles can exist in a part of the system, while in another part neighboring phases are highly correlated and thus the phase profile there looks like a smooth curve.
Such a picture in a lattice of identical oscillators, first presented and analyzed by KB~\cite{Kuramoto-Battogtokh-02}, is a typical chimera pattern.
However, for a theoretical description it is not appropriate to operate with non-smooth phase profiles, and thus a description based on the dynamics of the coarse-grained order parameters has been developed~\cite{Laing-09,Bordyugov-Pikovsky-Rosenblum-10}.
These order parameters, being defined as averages over small spatial domains, are per definition continuous, and one writes partial differential equations for them~\cite{laing2015chimeras,Smirnov-Osipov-Pikovsky-17}.
However, these equations are well-posed if the oscillators are not identical, but have a spread of natural frequencies.
Then equations for the order parameters (obtained usually using the Ott-Antonsen ansatz) contain damping terms that regularize the dynamics (another possible regularization is inclusion of viscosity~\cite{Smirnov_etal-22}).

In one-dimensional lattices, one typically considers a left{\,}--{\,}right symmetrical coupling between oscillators.
In this case, it is natural to expect chimera to stay (up to weak diffusion induced by finite-size fluctuations~\cite{Omelchenko_etal-10}), and such a spatial-temporal pattern indeed is observed in most setups.
However, in some cases traveling  solutions have been observed. 
In~\cite{Smirnov_etal-22}, traveling soliton chimera was studied.
It was demoinstrated that a directed motion is a finite-size effect, which disappears in the thermodynamic limit.
Most close to the topic of our paper are studies of traveling chimeras in~\cite{Xie_etal-14, Omelchenko-19}.  Xie et al.~\cite{Xie_etal-14} observed two types of traveling patterns in a system of identical units with symmetric in space coupling.
At some values of parameters traveling regular phase profiles in a lattice of identical phase oscillators were formed, moving with a slightly periodically modulated in time velocity. 
For other parameters, moving with a nearly constant velocity chimeras consisting of synchronous and asynchronous regions, have been observed.
Omelchenko~\cite{Omelchenko-19} considered non-identical units (so that the PDE approach based on the coarse-grained order parameter could be applied) with asymmetric coupling, and described different patterns and their stability.
However, his analysis could not be extended to the case of identical oscillators. 
We mention also that synchronization waves (moving patterns of different degrees of local synchrony) has been reported  in systems with local coupling~\cite{Smirnov_Osipov_Pikovsky-18} and with a combination of global and local coupling~\cite{dudkowski2019traveling}.

In this paper, we consider traveling chimera states in a system of identical oscillators. Our model is based on the KB setup,
with an additional advective term in the coupling. This model is introduced in Section~\ref{sec:bm}. Our basic observation 
is that a relatively smooth phase profile appears in such a system, which, however, can be well visualized for a large number
of oscillators only. This traveling regime is non-stationary and weakly irregular, and we illustrate it and describe
its statistical properties in Section~\ref{sec:il}. In Section \ref{sec:tw} we construct a family of regular traveling wave
profiles of the phases. However, only some of them are stable (and if yes, in a certain range of the advection parameter only).

\vspace{-1.25mm}
\section{Basic model: advection term in coupling of oscillators} \label{sec:bm}
In this section, we introduce the basic model which incorporates an advective term in the coupling of oscillators.
It is based on the famous KB setup~\cite{Kuramoto-Battogtokh-02}.  
The original KB model is formulated as a one-dimensional, periodic in space with period $1$ array of phase oscillators $\vp(x,t)$ coupled via a complex diffusive field $H(x,t)$. 
In the continuous in space formulation the equations read \vspace{-1.25mm}
\begin{subequations}
\begin{gather}
\frac{\partial\varphi}{\partial{t}}=\mathrm{Im}\!\left(H(x,t)e^{-i\vp(x,t)-i\alpha}\right), \label{eq:kb}\\
\tau\frac{\partial{H}}{\partial{t}}-\frac{\partial^{2}{H}}{\partial{x}^{2}}+\kappa^{2}H=-\kappa^{2}e^{i\vp(x,t)}. \label{eq:kbh} \vspace{-1.25mm}
\end{gather}
\label{eq:kbtwo}
\end{subequations}
Here, $\kappa^{-1}$ is the characteristic diffusion length of the local driving field $H(x,t)$, $\tau$ is its characteristic time scale, and $\alpha$ is the phase shift in the coupling.
Below, periodic boundary conditions $\vp(x+1,t)=\vp(x,t)$, $H(x+1,t)=H(x,t)$, and $\partial_{x}H(x+1,t)=\partial_{x}H(x,t)$ are assumed.
The coupling~\eqref{eq:kbh} is motivated by a chemical interpretation of the dynamics according to works of Y.~Kuramoto and co-workers~\cite{Kuramoto_etal-00, Kuramoto-Battogtokh-02, Tanaka-Kuramoto-03, Shima-Kuramoto-04}.
In this interpretation, lump oscillators interact via a diffusive medium.
We extend the setup \eqref{eq:kbtwo} by adding advection with velocity $2V$ to the evolution of the field $H$,  so that the second equation is now the advection-diffusion equation \vspace{-0.5mm}
\begin{equation}
\tau\frac{\partial{H}}{\partial{t}}+2V\frac{\partial{H}}{\partial{x}}-\frac{\partial^{2}{H}}{\partial{x}^{2}}+\kappa^{2}H=-\kappa^{2}e^{i\vp(x,t)}. \label{eq:difh} \vspace{-0.5mm}
\end{equation}
This makes the interaction between the oscillators asymmetric in space, so that traveling solutions are to be expected.

\begin{figure}[t]
\centering
\includegraphics[width=0.5\textwidth]{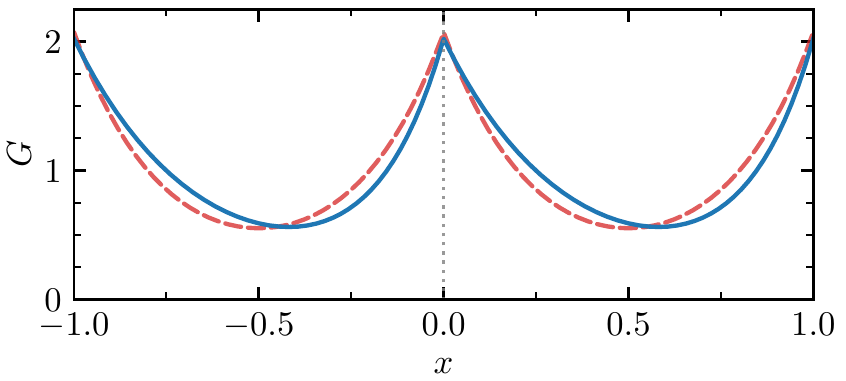} \vspace{-2.5mm}
\caption{The Green function (coupling kernel) for $\kappa=4$ and $V=0$ (red dashed curve), $V=1.25$ (blue solid curve).} \vspace{-2.5mm}
\label{fig:ker3}
\end{figure}

Following the original KB formulation~\cite{Kuramoto-Battogtokh-02}, we consider the case of very fast relaxation of field $H(x,t)$, i.e. the limit $\tau\!\to\!{0}$ (see \cite{Bolotov_etal-22, Smirnov_etal-22} for the analysis of a general situation $\tau\!>\!0$). 
In this case, the field $H(x,t)$ can be represented  via the Green function of the time-independent equation
$\bigl(d^{2}\!\bigl/dx^{2}\bigr.\!-\!2Vd\bigl/dx\bigr.\!-\!\kappa^{2}\bigr)G\!=\!-\kappa^{2}\delta(x)$ with periodic boundary conditions at $x=0,1$, which reads \vspace{-0.5mm}
\begin{equation} \label{eq:gf}
G(x)=\frac{\kappa^2}{2\sqrt{\kappa^2+V^2}}\left
( \frac{e^{\mu_2 x}}{e^{\mu_2}-1}-\frac{e^{\mu_1 x}}{e^{\mu_1}-1}\right),
\quad \mu_{1,2}=V\mp \sqrt{\kappa^2+V^2},\quad {0}\leq{x}\leq{1}. \vspace{-0.5mm}
\end{equation}
With this function, the phase dynamics according to~\eqref{eq:kb} can be written as an integral equation \vspace{0mm}
\begin{equation} \label{eq:egf}
\frac{\partial\vp(x,t)}{\partial{t}} =\int_{0}^{1}\!\!G(x-\tilde{x})\sin(\vp(\tilde{x},t)-\vp(x,t)-\alpha) d\tilde{x}. \vspace{0mm}
\end{equation}
We illustrate the Green function $G(x)$ in Fig.~\ref{fig:ker3}.

For performing numerical simulations, we discretize Eq.~\eqref{eq:egf} as follows.
One considers a finite set of $N$ oscillators at poisions $x_{n}=(n-1)\bigl/N\bigr.$, and approximates the integral as a sum.
As a result one has a system of $N$ ordinary differential equations for phases $\vp_{n}$: \vspace{-0.5mm}
\begin{equation} \label{eq:discr}
\frac{d\vp_{n}}{dt}=\frac{1}{N}\sum_{\tilde{n}=1}^{N}G(x_{n}-x_{\tilde{n}})\sin(\vp_{n}-\vp_{\tilde{n}}-\alpha).
\end{equation}

\vspace{-1.25mm}
\section{Traveling chimera and its properties} \label{sec:il}
\vspace{-1.25mm}
\subsection{Pictures of chimera}
\begin{figure}[t]
\centering
\includegraphics[width=\textwidth]{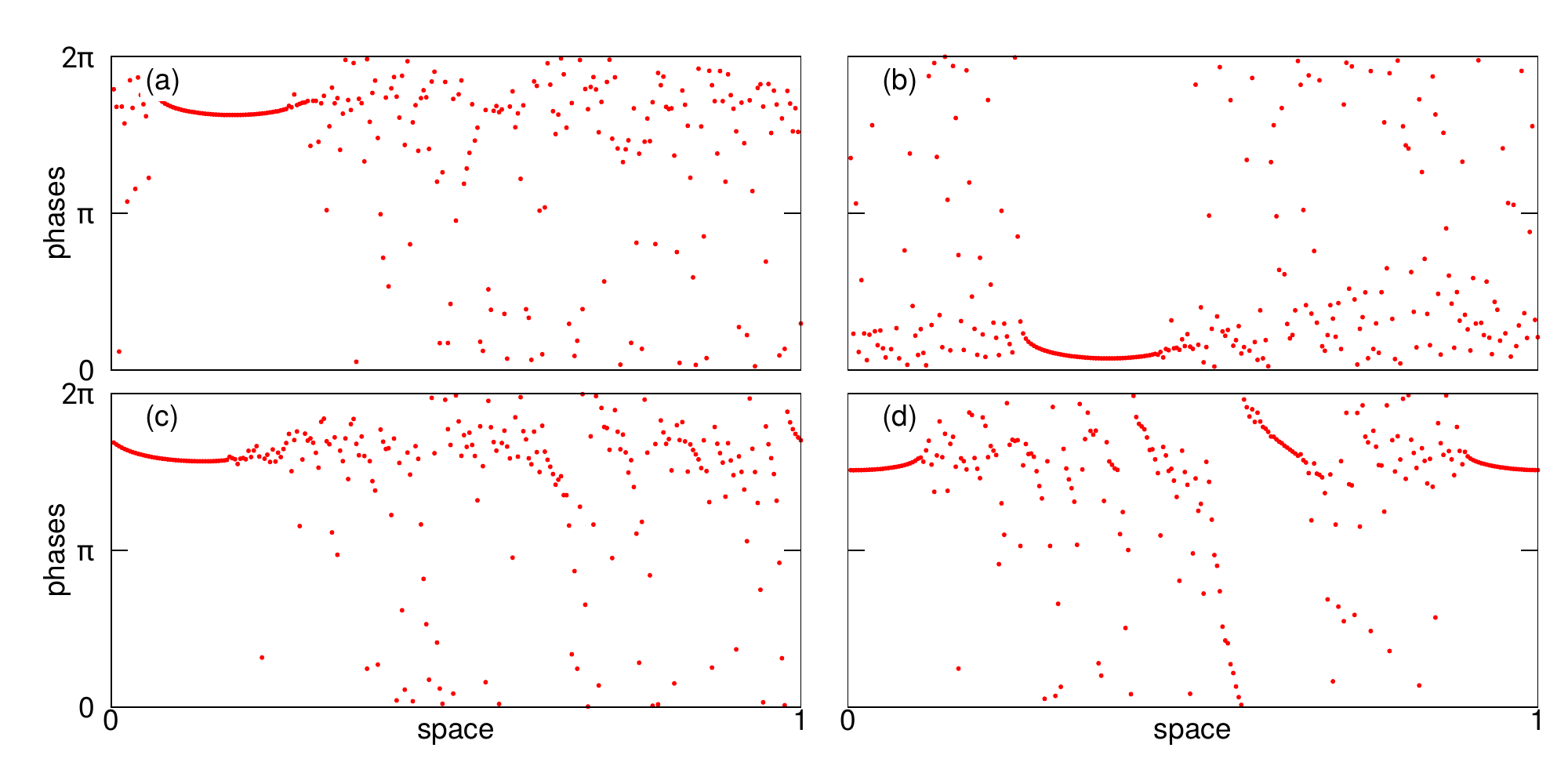} \vspace{-2.5mm}
\caption{Chimera states at different moments of time in a lattice of $N=256$ units. 
Panel (a): $t=300$, panel (b): $t=550$, panel (c): $t=1750$, panel (d): $t=3500$.
Parameters: $\alpha=1.5$, $\kappa=5$, $V=0.1$.} \vspace{-2.5mm}
\label{fig:c256}
\end{figure}

In this section, we present results of direct numerical simulations of model \eqref{eq:discr}.
We always start with a standing chimera pattern existing for the symmetric case $V=0$: the oscillators are synchronous in one spatial domain, and asynchronous in another one.
For $V\neq 0$, this state starts to move.
We illustrate what is observed in a system with a relatively small number of units in Fig.~\ref{fig:c256}. 
One can see that in snapshots (a,b,c) there is  one synchronous domain and one asynchronous domain, so that the moving chimera pattern looks rather similar to the standing chimera.
In panel (d) one can see an additional  synchronous region, however it does persist. 

\begin{figure}[t]
\centering
\includegraphics[width=\textwidth]{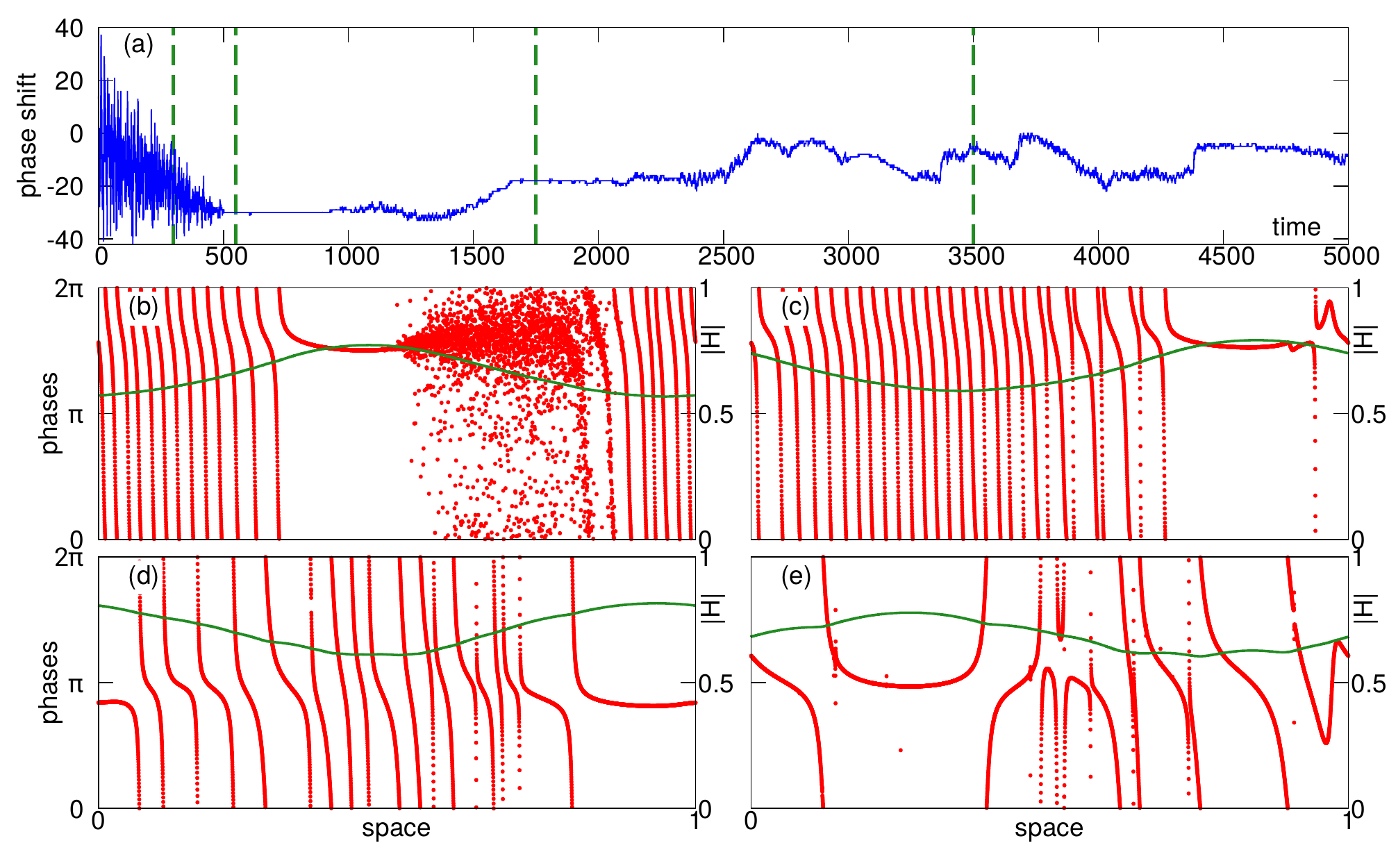} \vspace{-2.5mm}
\caption{Evolution of chimera for the same parameters as in Fig.~\ref{fig:c256}, but for $N=8192$.
Panels (b-e): snapshots at the times indicates by vertical dashed lines in panel (a).
Red dots: phases $\vp(x,t)$, green line: profile of the field $|H(x,t)|$ (right axis).
(Panel (a): time evolution of the spatial phase shift $M$ defined according to \eqref{eq:phsh}.} \vspace{-2.5mm}
\label{fig:c8192}
\end{figure}

In Fig.~\ref{fig:c256} the number of oscillators is $N=256$.
The picture is rather different if one considers a dense set of oscillators with $N=8192$.
In Fig.~\ref{fig:c8192} we present the evolution of an initial chimera at the same times and for the same parameters as in Fig.~\ref{fig:c256}. 
One can see that a continuous spatial profile $\vp(x,t)$ develops, without a disordered domain. 
At the first stage, the synchronous domain moves to the right, and behind it a continuous profile of the phases forms.
In panel (b) one can see a stage at which an initial strongly disordered domain still exists (the synchronous domain has been shifted by a distance less than $1$).
The strongly disordered domain disppears at $t\approx{500}$, and after that an ordered profile appears (panel (c)).
This profile is however unstable and a modulation develops: distances between some branches become smaller, and between some other branches become larger (this modulation is already seen in panel (c) of Fig.~\ref{fig:c8192}). 
In the course of this irregular modulation some branches merge and disappear.
We call this regime weak turbulence, because it, on one hand, is irregular in large, but on the other hand, locally it at most places looks like a continuous phase profile.
It is illustrated in panels (d,e) of Fig.~\ref{fig:c8192}.
In all snapshots we also show the profiles of the driving field $|H(x,t)|$.
These profiles are rather smooth in all cases, because of diffusion.
The maximum position of the field $|H(x,t)|$ is in the mostly synchronous domain, where the phases are close to each other (they form a horizontal bar).

\vspace{-1.25mm}
\subsection{Phase shift number}
Next, we present characterizations of the phase profiles of moving chimera. 
An inspection of the panels (c,d) in Fig.~\ref{fig:c8192} reveals rather smooth in space profiles  
of phases. These profiles are wrapped to the interval $0\leq\varphi<2\pi$, but one can unwrap them
to the phases belonging to the real line. It is thus possible to characterize them
with the ``spatial rotation number'', we will call it the phase shift. 
We define the total phase shift along the circular spatial domain $0\leq x <1$ as \vspace{-1.25mm}
\begin{equation} \label{eq:phsh}
M=\frac{1}{2\pi}\sum_{n=1}^{N-1}\!\mathrm{arg}\!\left[e^{i(\varphi_{n+1}-\varphi_{n})}\right]. \vspace{-1.25mm}
\end{equation}
One can see that this definition is invariant to shifts of the phases $\varphi_{n}\to\varphi_{n}\pm{2\pi}$, provided
that we stick to the definition of the $\mathrm{arg}$ function as $-\pi<\mathrm{arg}[z]\leq \pi$. 
Thus for smooth profiles $\Bigl|\mathrm{arg}\Bigl[e^{i(\varphi_{n+1}-\varphi_{n})}\Bigr]\Bigr|\ll{1}$
and the phase shift $M$ is defined properly. 
We will apply definition \eqref{eq:phsh} also to erratic 
profiles, where the phase shift along the spacial domain cannot be defined unambiguously.
The results of this analysis are presented in panel (a) of Fig.~\ref{fig:c8192}.
One can see that at the initial stage, where the smooth phase profile is still in the formation,
the phase shift strongly fluctuates in time; this is a clear indication for the intrinsic
non-smoothness of the phase profile (panel (b)). This stage finishes at $t\approx 500$, here the
smooth phase profile like in panel (c) is formed. One can see that quite for a long time the value of 
the phase shift is nearly a constant $M(t)\approx -30$. This state is however weakly unstable
and in the course of evolution for $t>920$ the value of $M$ changes significantly. Moreover, there
are visible fluctuations on a small time scale in the dependence $M(t)$, indicating that there are 
non-smooth changes of the 
phase (such small non-smooth domains are clearly seen in the snapshot panel (e)).

\vspace{-2.5mm}
\subsection{Velocity and lifetime}
\begin{figure}[t]
\centering
\includegraphics[width=0.625\columnwidth]{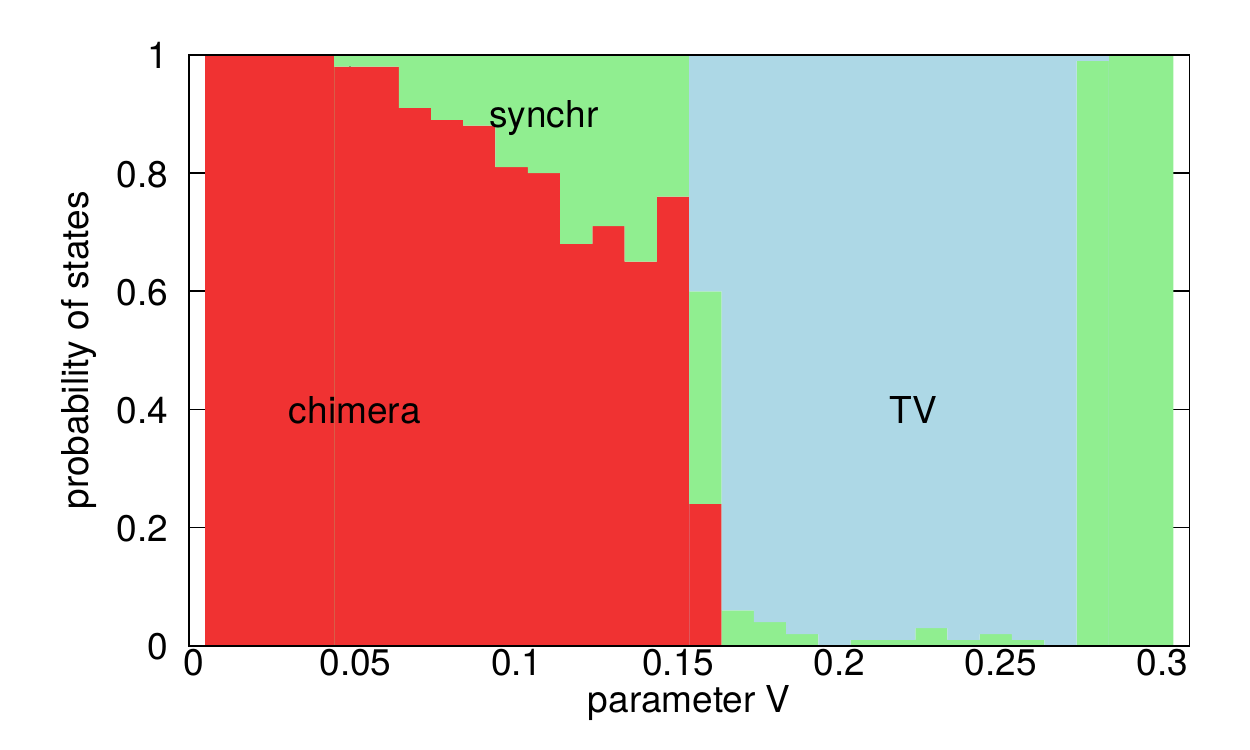} \vspace{-3.75mm}
\caption{A stacked plot of probability (obtained from 100 independent runs)
of different states after evolution of initial chimera up to time $T=5\cdot 10^5$.
Parameters: $N=2048$, $\alpha=1.5$, $\kappa=5$. We distinguish here 3 regimes: chimera like in  Fig.~\ref{fig:c8192} (red);
a synchronous regime with a linear or a constant phase profile (green) and a traveling wave (blue).} \vspace{-3.75mm}
\label{fig:lt}
\end{figure}

Above in Fig.~\ref{fig:c8192} we illustrated the evolution of traveling chimera for relatively short
time intervals. Numerical simulations on longer time intervals show, that for many parameter values
the described in   Fig.~\ref{fig:c8192} regime is a long transient, after which a regular state appears.
This regular state can be either a fully synchronous state where all the phases are equal, or a twisted wave where
all the phases build a linear in space profile, or a nontrivial regular traveling wave, to be discussed
in details in Section~\ref{sec:tw}. In Fig.~\ref{fig:lt} we report a statistical evaluation of the fate
of an initial chimera after a long time interval $T=5\cdot 10^5$, in dependence on parameter $V$. 
One can see that chimera always survives
for small $V$, while in the range $0.05\lesssim V\lesssim 0.15$ up to 35\% of all runs lead to synchrony.
Characteristic lifetimes of chimera in the latter cases are $\approx 2.5\cdot 10^5$ (we remind here that the simulations stopped
at $T=5\cdot 10^5$). In the range $0.16\lesssim V\lesssim 0.28$ the dominant asymptotic regime 
is a traveling wave,  the characteristic transient time from chimera to this state is $7\cdot 10^3\lesssim T_{tr}\lesssim 7\cdot 10^4$. At larger values of parameter $V\gtrsim 0.28$ a synchronous state arises after a relatively short 
characteristic transient time $T_{tr}\sim 5\cdot 10^3$.

\begin{figure}[!ht]
\centering
\includegraphics[width=0.625\columnwidth]{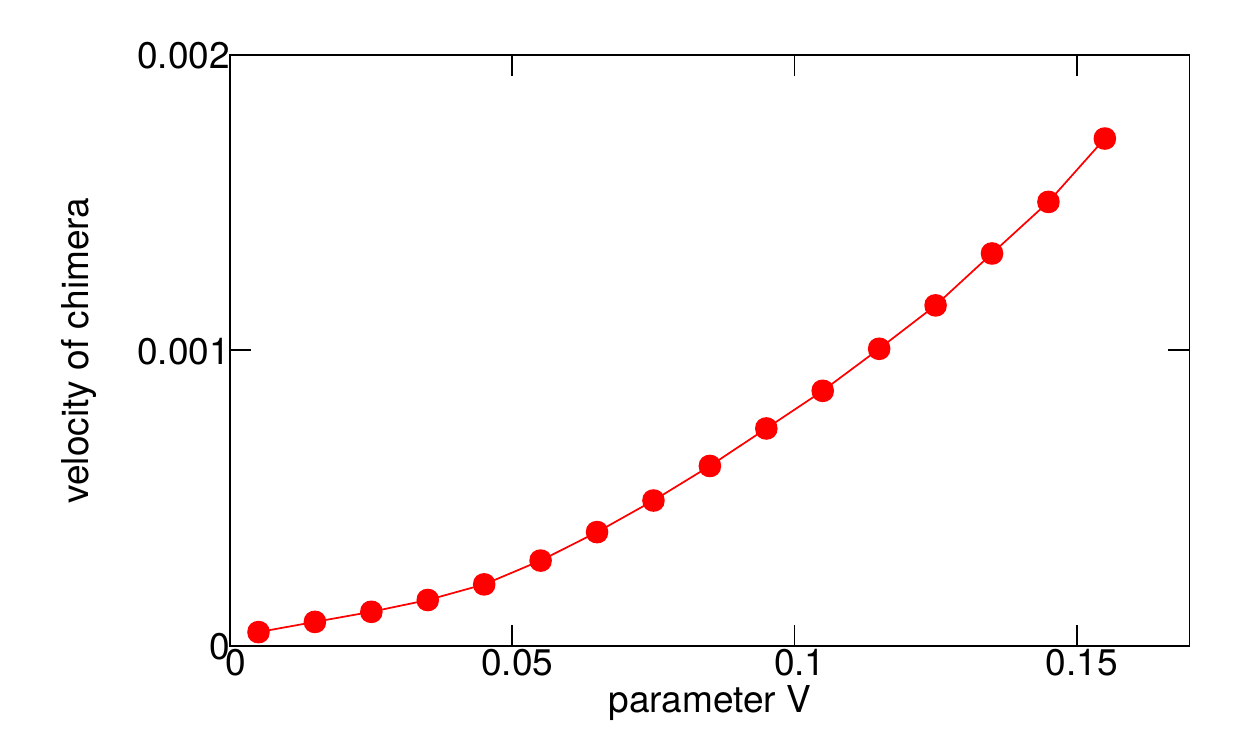} \vspace{-3.75mm}
\caption{Velocity of chimera vs advection parameter $V$ for $\kappa=5$, $\alpha=1.5$.} \vspace{-3.75mm}
\label{fig:vel}
\end{figure}

\begin{figure}[t]
\centering
\includegraphics[width=0.625\columnwidth]{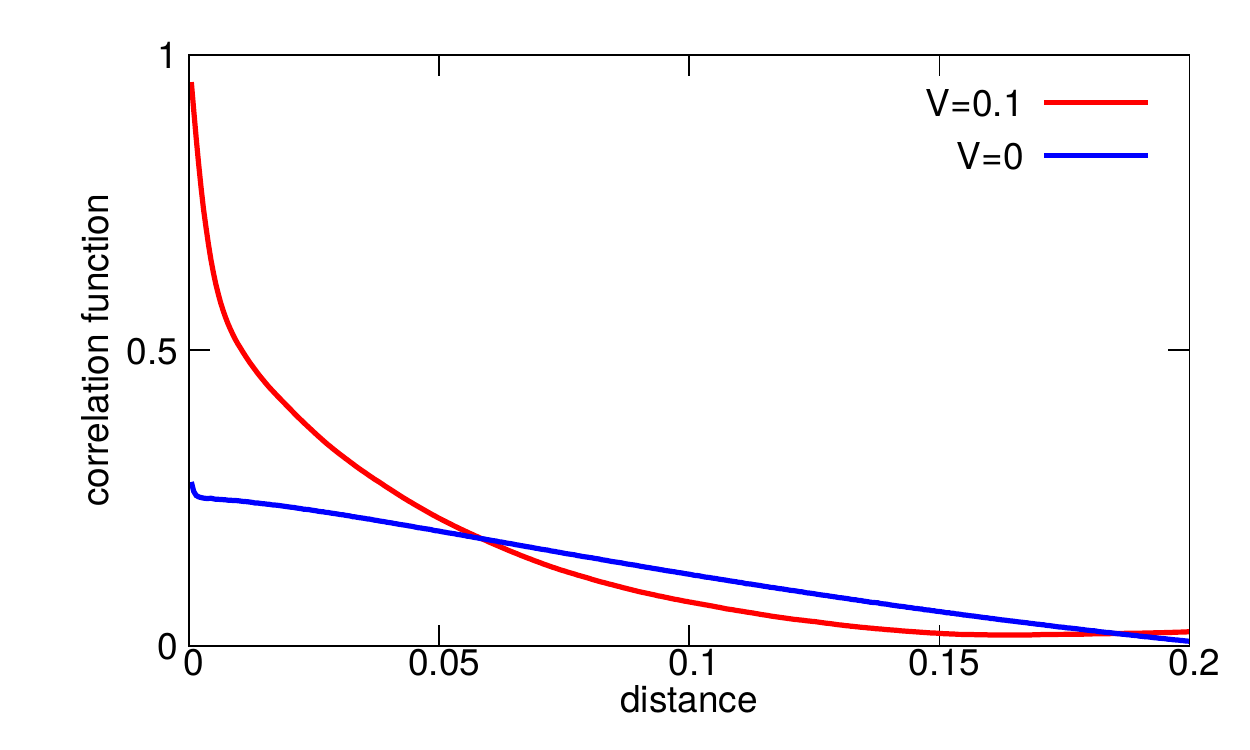} \vspace{-3.75mm}
\caption{Correlation function for $N=2048$ and two values of parameter $V$. Other parameters: $\alpha=1.5$, $\kappa=5$.} \vspace{-5mm}
\label{fig:corr}
\end{figure}

\looseness=-1
The observed situations where traveling chimera states are long transients should be juxtaposed with similar
observartions for standard chimeras (at $V=0$). According to Ref.~\cite{Wolfrum_Omelchenko-11}, standard chimeras 
are also long transients
evolving eventually to synchronous regimes. However, there the transition time grows exponentially with the number of units $N$,
so that for typical parameters no transition is observed for $N\gtrsim 50$. In the case of traveling chimeras above, 
we have not
found any significant dependence of the lifetime on the number of units $N$. We attribute this to the structure of the
phase profile, which is strongly correlated at small distances between the elements (see Fig.~\ref{fig:corr} below for quantitative characterization
of these correlations). Thus, the number of independent patches in the turbulent state can be esctimated as
the phase shift number $|M|$ (see Eq.~\eqref{eq:phsh}). Because this number only weakly depends on $N$ and is relatively small,
effective fluctuations leading eventually to a transition to a regular regime do not decrease with the number of units $N$.
This explains finite lifetimes even for systems with a large number of units.

Next, we discuss statistical properties of turbulent states.
We show the mean velocity of the chimeras in dependence on parameter $V$ in Fig.~\ref{fig:vel}.
This quantity was determined numerically according to the position of the maximum of the acting field $|H|$.
Indeed, this field, because of diffusion, is rather smooth, and at each moment of time it has a spatial profile with one maximum (see profiles in Fig.~\ref{fig:c8192}).
To reduce fluctuations, we calculated the position of the maximum as the phase of the complex spatial mode $\int_{0}^{1}|H(x,t)|e^{i2\pi{x}}dx$.
Remarkably, the velocity is proportional to the square of the advection parameter $v\sim V^2$.

\begin{figure}[!ht]
\centering
\includegraphics[width=0.625\columnwidth]{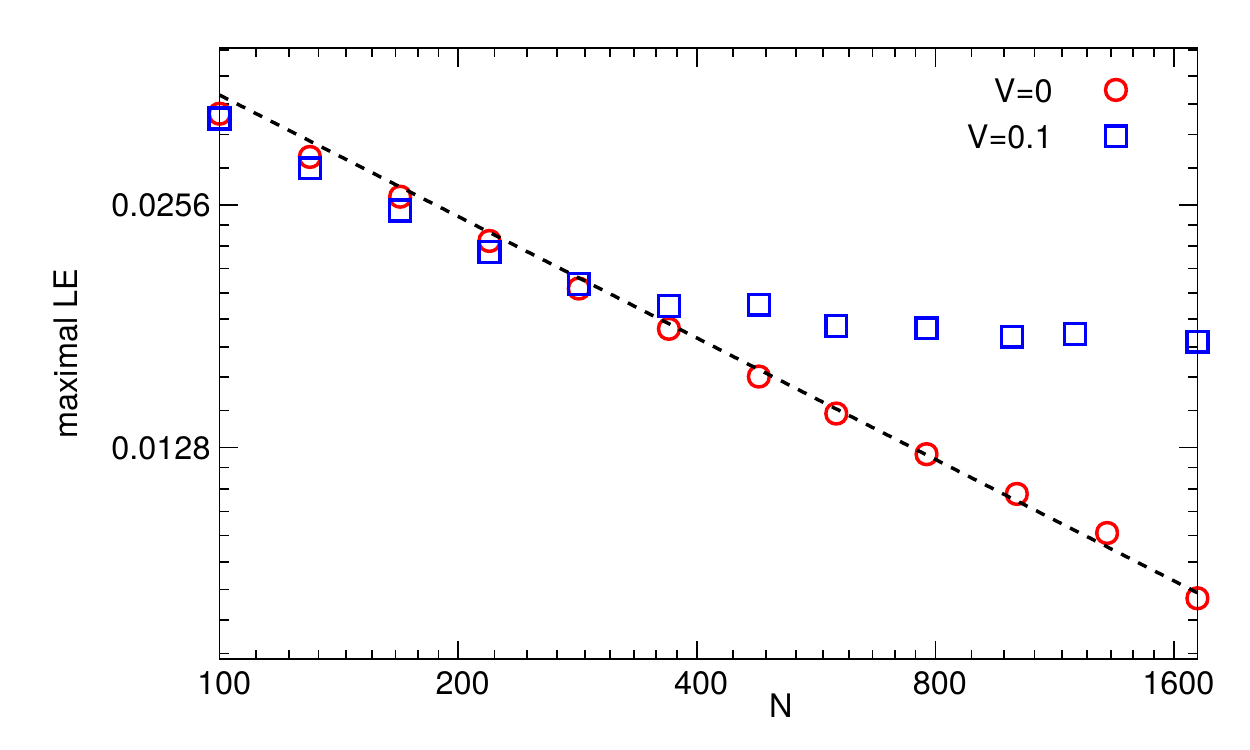} \vspace{-3.75mm}
\caption{Largest Lyapunov exponents for the standard standing chimera ($V=0$) and for traveling
turbulent chimera in dependence on the number of units $N$. The dashed line is the power law
$\sim N^{-1/2}$. Parameters: $\alpha=1.5$, $\kappa=5$.} \vspace{-3.75mm}
\label{fig:lyap}
\end{figure}

\vspace{-2.5mm}
\subsection{Cross-correlations}
One characteristic feature of traveling chimeras is that they have continuous (or at least with large domains of continuity) profiles of phases (if $N$ is large). 
To characterize this quantitatively, we calculated the cross-correlation function of the phases.
Because the phase are distributed non uniformly, it is appropriate to use a transformation to nearly uniformly distributed phases $\vp\to\theta$.
For this, the global order parameter is calculated $Z=\langle e^{i\vp}\rangle$, and a M\"{o}bius transform is performed 
\begin{equation}
e^{i\theta_{n}}=\frac{e^{i\vp_{n}}-Z}{1-Z^{\ast}e^{i\vp_{n}}}
\end{equation}
After this the quantity
\begin{equation}
\gamma(n)=\left\langle e^{i(\theta_{m}-\theta_{m+n})}\right\rangle
\end{equation}
is calculated. The correlation function shown in Fig.~\ref{fig:corr} is $|\gamma(n)|$ vs $\Delta x=n/N$.
One can see that for $V=0$ (i.e. for standing usual chimera) 
the correlations tend to $\approx 1/3$ for small distances $\Delta x$. This reflects
an average of full correlations in the synchronous domain and of absence of correlations in the 
asynchronous domain. In contradistinction, the correlation function for $V=0.1$
tends to
one at small distances, what indicates 
for continuity of the phase profiles.

\vspace{-1.25mm}
\subsection{Lyapunov exponents}
Next, we present calculations of the largest Lyapunov exponent of chimera states. The standard chimera (at $V=0$)
is known \cite{Omelchenko_etal-10,Wolfrum_Omelchenko-11} to be weakly chaotic, the largest Lyapunov exponent is positive.
However, chaos decreases with $N$ and disappears in the limit $N\to\infty$. In the thermodynamic limit, the field $H(x,t)$ acting on oscillators is stationary (in the corersponding reference frame), and the Lyapunov exponents of particular oscillators are
either negative (in the synchronous domain) or zero (in  the asynchronous domain). In the traveling turbulent chimera we observe
a different behavior of the largest Lyapunov exponent (Fig.~\ref{fig:lyap}) in dependence on the number of the oscillators
$N$. It  decreases for small $N$, but than saturates.
This level of the largest Lyapunov exponent characterizes chaoticity of the turbulent state, 
which exists apparently
also in the thermodynamic limit $N\to\infty$.

\vspace{-1.25mm}
\section{Traveling wave solutions} \label{sec:tw}
\vspace{-1.25mm}
\subsection{Equations for traveling wave solutions} \label{sec:eqtw}
In this section, we focus on the traveling waves in form of a continuous phase profile with a fixed shape, moving at a constant velocity.
Our starting point is the KB setup, formulated as integro-differential equation~\eqref{eq:gf} with an asymmetric in space exponential-type kernel~\eqref{eq:egf}.
To formulate tractable equations for the traveling waves, we employ full equivalence between the modified KB model under consideration and an oscillator medium closed in a ring consisting of identical elements interacting via a rapidly relaxing advection-diffusion mean field.
This means, we use the representation whereby the dynamics of the phase $\varphi(x,t)$ of each unit is given by Eq.~\eqref{eq:kb}, where the instantaneous distribution of complex valued coupling field $H(x,t)$ is governed by Eq.~\eqref{eq:difh} (in which we set $\tau=0$ due to our assumption of fast relaxation process) with periodic boundary conditions.

We apply the following traveling wave ansatz:
\begin{equation} \label{eq:trwave}
\varphi(x,t)=\Omega{t}+\phi(\xi), \quad H(x,t)=h(\xi)e^{i\Omega{t}}, \quad \xi=x-vt \quad (v\neq{0}),
\end{equation}
where $\Omega$ is the (unknown) rotation frequency and $v$ is the (unknown) velocity of the corresponding wave pattern.
Both of these unknown constants play the role of the two unique parameters of the moving structure repeatedly running over the system, and are to be determined simultaneously with a continuous phase profile $\phi(\xi)$ and an inhomogeneous profile $h(\xi)$ of the self-consistent acting field.
Substituting~\eqref{eq:trwave} in  Eqs.~\eqref{eq:kb} and~\eqref{eq:difh} with $\tau=0$, we obtain a 5-dimensional system of ordinary differential equations (ODEs) (because $h(\xi)$ is complex):
\begin{subequations}
\begin{gather}
\frac{d{\phi}}{d{\xi}}=\frac{1}{v}\biggl[\Omega-\mathrm{Im}\!\left(h(\xi) e^{-i\phi(\xi)-i\alpha}\right)\biggr],  \label{eq:eqs-qph-qp} \\
\frac{d^{2}{h}}{d{\xi}^{2}}-2V\frac{d{h}}{d{\xi}}-\kappa^2 h=\kappa^2 e^{i\phi(\xi)}. \label{eq:eqs-qph-h}
\end{gather}
\label{eq:eqs-qph}
\end{subequations}
Because we consider an ensemble of nonlocally coupled phase oscillators on a ring with unit length (in dimensionless variables), 
 functions $\phi(\xi)$ and $h(\xi)$ satisfy the following periodicity conditions at the ends of the interval ${0}\leq\xi\leq{1}$:
\begin{equation} \label{eq:pbc-qph}
\phi(\xi=0)=\phi(\xi=1)-2\pi{M}, \quad h(\xi=1)=h(\xi=0), \quad h'(\xi=1)=h'(\xi=0).
\end{equation}
Hereafter primes at functions denote derivatives with respect to the traveling coordinate $\xi$.
Noteworthy, in~\eqref{eq:pbc-qph} we take into account that, with bypass over the full spatial domain, the phase determined in the thermodynamic limit at each point of the oscillatory media can make several rotations by $2\pi$ (cf. Fig. \ref{fig:c8192}(c)).
The number of these rotations is a topological characteristics of a traveling wave characterized by an additional integer parameter $M$  (cf. Eq.~\eqref{eq:phsh}).

\vspace{-1.25mm}
\subsection{Procedure of finding traveling waves solutions}
Here we describe the adopted procedure for finding solutions of the system~\eqref{eq:eqs-qph},~\eqref{eq:pbc-qph}.
Because of the phase shift invariance ${\varphi}\to{\varphi+\varphi_{0}}$  and the space shift invariance  ${x}\to{x+x_{0}}$  of the KB model (where $\varphi_{0}$ and $x_{0}$ are arbitrary constants), one can assume (without loss of generality) that such solutions satisfy the following two equalities: $\mathrm{Im}\bigl[h(\xi=0)\bigr]\!=0$ and $\mathrm{Re}\bigl[h'(\xi=0)\bigr]\!=0$.
The first of them means that the phase of the complex field $h(\xi)$ can be set to zero at the origin $\xi=0$, and the second means that the smooth distribution of the absolute value of $h(\xi)$ has an extremum at the origin $\xi=0$ of the moving coordinate system.

Therefore, one has 5 unknown quantities $\phi(\xi=0)=\mathcal{Q}$, $\mathrm{Re}\bigl[h(\xi=0)\bigr]=\mathcal{R}$, $\mathrm{Im}\bigl[h'(\xi=0)\bigr]=\mathcal{S}$, $\Omega$, $v$ and 5 periodicity conditions~\eqref{eq:pbc-qph} to be fulfilled.
Actually, we arrive at the system of nonlinear equations for the announced above real variables $\mathcal{Q}$, $\mathcal{R}$, $\mathcal{S}$, $\Omega$, and $v$:
\begin{equation} \label{eq:nleqs}
\begin{gathered}
\phi(\xi=1|\mathcal{Q,R,S},\Omega,v)-\mathcal{Q}-2{\pi}M=0, \\
\mathrm{Re}\bigl[h(\xi=1|\mathcal{Q,R,S},\Omega,v)\bigr]-\mathcal{R}=0, \quad
\mathrm{Im}\bigl[h(\xi=1|\mathcal{Q,R,S},\Omega,v)\bigr]=0, \\
\mathrm{Re}\bigl[h'(\xi=1|\mathcal{Q,R,S},\Omega,v)\bigr]=0, \quad
\mathrm{Im}\bigl[h'(\xi=1|\mathcal{Q,R,S},\Omega,v)\bigr]-\mathcal{S}=0, \\
\end{gathered}
\end{equation}
where the real function $\phi(\xi|\mathcal{Q,R,S},\Omega,v)$ and the complex field $h(\xi|\mathcal{Q,R,S},\Omega,v)$  together with its derivative $h'(\xi|\mathcal{Q,R,S},\Omega,v)$ for a given values of $\Omega$ and $v$ describe the trajectory 
in the phase space of the 5-dimensional system of ODEs~\eqref{eq:eqs-qph} beginning at the initial point $(\mathcal{Q},\mathcal{R},0,0,\mathcal{S})$.

As a result, for fixed values of the parameters $\alpha$, $\gamma$, and $\kappa$, the problem of finding a traveling wave reduces to the problem of finding roots of Eqs.~\eqref{eq:nleqs} and can be solved numerically (with high precision) by a so-called shooting procedure based on the Newton-Raphson method.
The corresponding approach has a sufficiently good convergence rate, but can have  a rather small convergence domain, which is the main stumbling block for the numerical calculations of a fixed point of the system~\eqref{eq:nleqs} in this way.
Thus, in order to avoid such an inherent impediment, it is necessary to find a good initial approximation close to the genuine values of $\mathcal{Q}$, $\mathcal{R}$, $\mathcal{S}$, $\Omega$, and $v$.

\vspace{-1.25mm}
\subsection{Construction of an initial approximation}
If one already has a solution of Eqs.~\eqref{eq:nleqs} or its approximation for some parameters $\alpha$, $\gamma$, and $\kappa$, then a continuation 
for neighboring values of parameters is an obvious straightforward approach, in which at each small step the  Newton-Raphson to numerically find a fixed point of the system~\eqref{eq:nleqs} can be implemented.
Substantially, the strategy is to start from a case where the phase profile moving at a constant velocity with a permanent shape is roughly known (hence, one can get adequate estimations for the genuine values of $\mathcal{Q}$, $\mathcal{R}$, $\mathcal{S}$, $\Omega$, and $v$) and to change parameters gradually to remain within the convergence domain of the Newton-Raphson method.
Because parameter $M$ is discrete, one cannot continue solution in it, thus one has to find at least one good estimation for each desired value of $M$. Practically, we perform continuation in the advection parameter $V$.

In order to obtain a preliminary information about nonuniform phase profiles, spatial structure of acting mean field and properties (including estimations for values of a rotation frequency $\Omega$ and a propagation velocity $v$) of at least one example of a traveling solution for a given set of parameters $\alpha$, $\gamma$, $\kappa$, and $M$, we develop an auxiliary iterative procedure. 
We successfully implement and approve this approach, but with no any rigorous mathematical proof of convergence.
Actually, we confirm efficiency and performance of the developed method only experimentally by direct numerical calculations for a number of different cases.
However, such evidence is sufficient for the practical purposes of searching for starting points for the shooting procedure and further constructing families of traveling wave patterns using the control-parameter continuation concept.
The main idea of our auxiliary approach is as follows.

The complex field $h(\xi)$ is a periodic function of spatial coordinate $\xi$ with unit period.
Supposing such a dependence $h(\xi)$ and assuming that parameters $\Omega$ and $v$ are approximately known (saying, from priori information), at the iteration, we 
first fix  $h(\xi),\Omega,v$ and consider them to be independent of the phase profile $\phi(\xi)$. Then, we find the profile $\phi(\xi)$ for given  $h(\xi),\Omega,v$ as follows.
Substituting the current approximation for $h(\xi)$ to Eq.~\eqref{eq:eqs-qph-qp}, one gets an equation that can be interpreted as 
the Adler equation with a periodic forcing in the nonlinear term.
Transforming to a variable $z=e^{i(\phi+\alpha)}$, Eq.~\eqref{eq:eqs-qph-qp} can be written in a complex form
\begin{equation} \label{eq:riccati}
\frac{dz}{d\xi}=\frac{1}{2v}\!\left(h^{\ast}\!(\xi)z^{2}+2i\Omega{z}-h(\xi)\right),
\end{equation}
which is the complex Riccati equation with periodic coefficients~\cite{Omel-BrChimeras-2022, Marvel-2009, Wilczynski-2008, Campos-1997}.
Here and below, symbol $^{\ast}$ denotes complex conjugate.
Noteworthy, this equation is partially reminiscent of the Ott-Antonsen equation for a coarse-grained complex order parameter 
(e.{\,}g., see~\cite{laing2015chimeras, Smirnov-Osipov-Pikovsky-17, Smirnov_etal-22, Omel-BrChimeras-2022}).
Formally, Eq.~\eqref{eq:riccati} can be considered not only on the unit circle $|z|=1$ but also inside the unit disc $|z|<1$ of the complex plane.
In the paper~\cite{Omel-BrChimeras-2022}, it is shown that, in general, there exists a unique stable solution to Eq.~\eqref{eq:riccati} starting from the initial condition $z(\xi=0)=z_{0}$, where $|z_{0}|\leq{1}$, and lying entirely in such a closure of the corresponding domain.
Moreover, if $|z_{0}|<1$ or $|z_{0}|=1$, then $|z(\xi)|<1$ or $|z(\xi)|=1$ for all $\xi>0$, respectively, i.{\,}e. every solution $z(\xi)$ of Eq.~\eqref{eq:riccati} satisfying $|z_{0}|<1$ remains trapped inside the domain $|z|<1$, and each trajectory starting from the initial point $|z_{0}|={1}$ on the unit circle of the complex plane stays on the given circle.

\begin{figure}[t]
\centering\includegraphics[width=\textwidth]{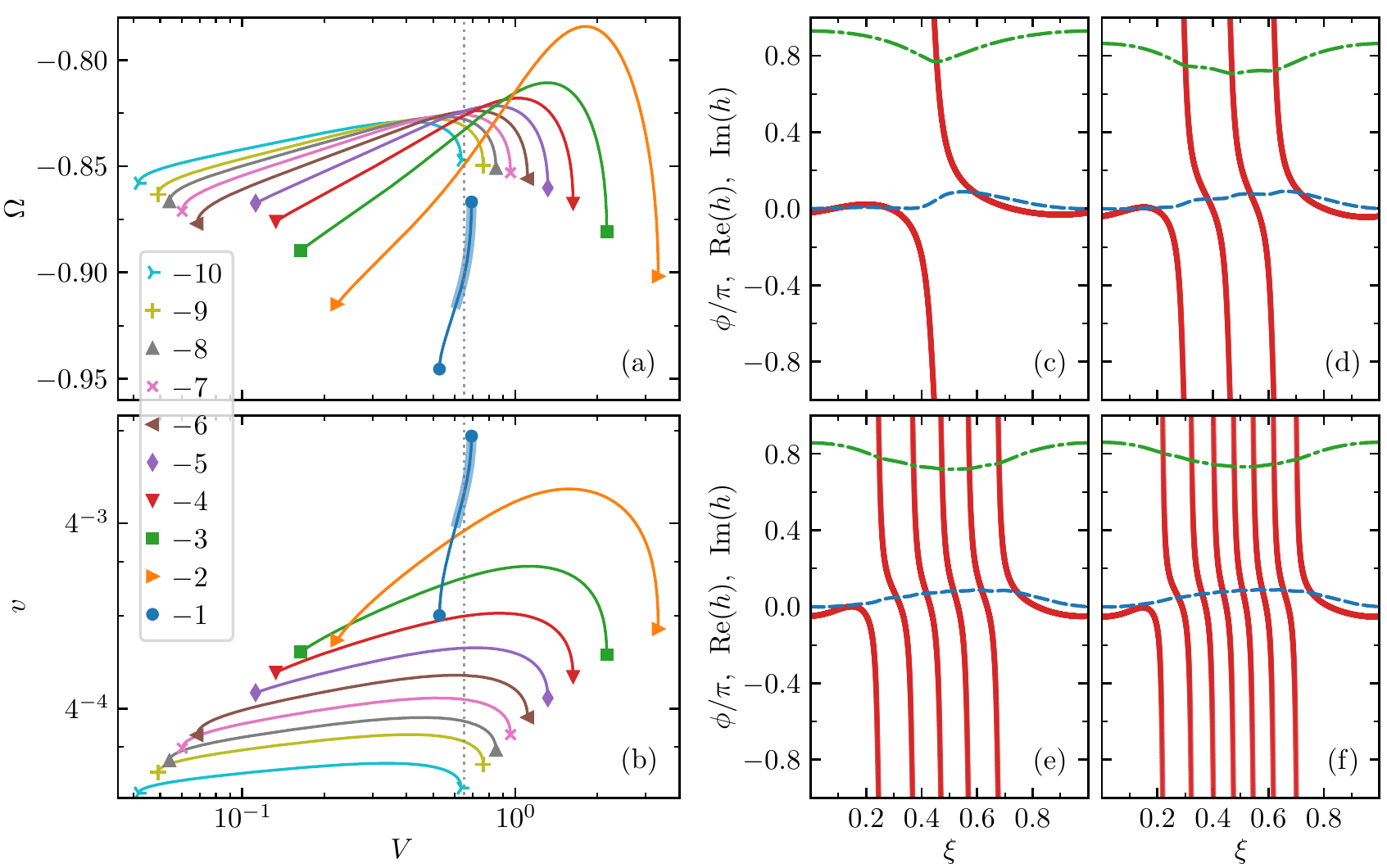} \vspace{-2.5mm}
\caption{(a),{\,}(b) Traveling wave solutions: dependence of frequency $\Omega$ and wave velocity $v$ on advection parameter $V$ for $M=-1,-2,\ldots,-10$.
Each line connecting two identical markers corresponds to a certain value of the parameter $M$ (see the graphic legend from bottom to top).
Here, we also depicted the stability of phase profiles moving at constant velocities.
Unstable solutions are shown with a thin part of the respective curve, stable solutions with a thickened part of the curve.
In this case, only the traveling patterns with $M=-1$ are stable only. 
(c){\,}--{\,}(f) Examples of traveling wave solutions for $V=0.65$ (dotted vertical lines on panels (a) and (b)) and several values of $M$: (c) $M=-1$, (d) $M=-3$, (e) $M=-5$, and (f) $M=-7$. 
The dotted red line merging into a solid curve is the phase profile, the dash-dot green line is the real part of the complex mean field $h(\xi)$, and the dashed blue line depicts its imaginary part.
Other parameters are the same as the parameter values used in the original KB article~\cite{Kuramoto-Battogtokh-02}: $\kappa=4$, $\alpha=1.457$. \vspace{-2.5mm}}
\label{fig:branches-1}
\end{figure}

It is well-known (e.{\,}g., see~\cite{Omel-BrChimeras-2022, Marvel-2009, Wilczynski-2008, Campos-1997}) that the Poincar\'{e} map of the periodic complex Riccati equation coincides with the M\"{o}bius transformation.
In our case, this M\"{o}bius transformation maps the closed unit disk $|z|\leq{1}$ onto itself, thus it can be written in the canonical form~\cite{Gong-2020, Marvel-2009}
\begin{equation} \label{eq:mobius}
\mathcal{M}_{q,\psi}(z)=\frac{q+e^{i\psi}z}{1+q^{\ast}e^{i\psi}z},
\end{equation}
with parameters $q$ and $\psi$, where $q$ is on the open complex unit disc $|q|<1$, and $e^{i\psi}$ on the complex unit circle.
Here, we use the same parametrization as in Refs.~\cite{Gong-2020, Marvel-2009}, where one can find properties of the M\"{o}bius transformation~\eqref{eq:mobius} and justifications that the Poincar\'{e} map of Eq.~\eqref{eq:riccati} with periodic coefficients $h(\xi)$ and $h^{\ast}(\xi)$ is described by formula~\eqref{eq:mobius}.
Note, the transformation $\mathcal{M}_{q,\psi}(z)$ can be applied to any complex number $z$ in the complex domain $|z|\leq{1}$ and leaves this domain invariant (as we need).
According to~\cite{Omel-BrChimeras-2022}, the values of two parameters $q$ and $\psi$ in~\eqref{eq:mobius} can be uniquely determined using two solutions $\mathcal{Z}_{0}(\xi)$ and $\mathcal{Z}_{1}(\xi)$ of Eq.~\eqref{eq:riccati} starting from the initial conditions $\mathcal{Z}_{0}(\xi=0)=0$ and $\mathcal{Z}_{0}(\xi=0)=1$, respectively.
Evaluating $\mathcal{Z}_{0}(\xi=1)=\zeta$ and $\mathcal{Z}_{1}(\xi=1)=e^{i\vartheta}$ by direct numerical calculations of the corresponding initial value problems for the complex Riccati equation~\eqref{eq:riccati} on the period of the function $h(\xi)$ and implying the definition~\eqref{eq:mobius} of the related Poincar\'{e} map, we obtain the following expressions for $q$ and $e^{i\psi}$:
\begin{equation} \label{eq:q-psi}
q=\zeta, \quad e^{i\psi}=\frac{\zeta-e^{i\vartheta}}{\zeta^{\ast}e^{i\vartheta}-1}.
\end{equation}
It is worth mentioning that, in our case, the inequality $|q|=|\zeta|<1$ is always satisfied.

Then, every periodic solution of Eq.~\eqref{eq:eqs-qph-qp} corresponds to a fixed point $\bar{z}$ of the Poincar\'{e} map coinciding with the M\"{o}bius transformation~\eqref{eq:mobius}.
In other words, to find a closed path of Eq.~\eqref{eq:eqs-qph-qp} for a given periodic complex function $h(\xi)$, we need to find
a fixed point of transformation~\eqref{eq:mobius} with constant map parameters $q$ and $\psi$ determined by expressions~\eqref{eq:q-psi}.
Consequently, we arrive at the following quadratic equation: 
\begin{equation}
q^{\ast}\bar{z}^{2}-\bigl(1-e^{-i\psi}\bigr)\bar{z}-qe^{-i\psi}=0,
\end{equation}
which has, in general, two roots $\bar{z}_{1}$ and $\bar{z}_{2}$ with the properties $\bar{z}_{1}+\bar{z}_{2}=\bigl.\bigl(1-e^{-i\psi}\bigr)e^{i\varsigma}\bigr/\varrho${\,},~ and ~$\bar{z}_{1}\bar{z}_{2}=e^{i(2\varsigma+\pi-\psi)}$, where $\varrho$ ($0\leq\varrho<1$) and $\varsigma$ are the amplitude and phase of the complex value $q=\varrho{e^{i\varsigma}}$, respectively.
These properties allow one to define fixed points of the M\"{o}bius map that are of interest to us.
In particular, from the second relation for $\bar{z}_{1}$ and $\bar{z}_{2}$ it follows that either the two fixed points are on the unit circle of a complex plane, or one of them is inside and the other outside the unit circle.
For us only the former case is relevant, since we are looking for traveling wave patterns 
with a strongly inhomogeneous but continuous phase profile.
Such a profile is given by Eq.~\eqref{eq:eqs-qph-qp} to which one can transform the periodic complex Riccati equation~\eqref{eq:riccati}  on the manifold $|z(\xi)|=1$.
In this case, for the two fixed points we obtain the following expressions: $\bar{z}_{1,2}=e^{i(\Psi\pm\Theta)}$, where $\Psi=\varsigma+(\pi-\psi)\!\left/2\right.$, and $\Theta$ is determined by equality $\varrho\cos\Theta=\sin\!\left(\psi\!\left/2\right.\right)$.
Thus, the condition for the two fixed points on the unit circle is $\left|\sin\!\left(\psi\!\left/2\right.\right)\right|\leq\varrho$.
Noteworthy, the fulfillment of this condition is to be checked at each step of the auxiliary iterative procedure we develop.

\begin{figure}[t]
\centering
\includegraphics[width=\textwidth]{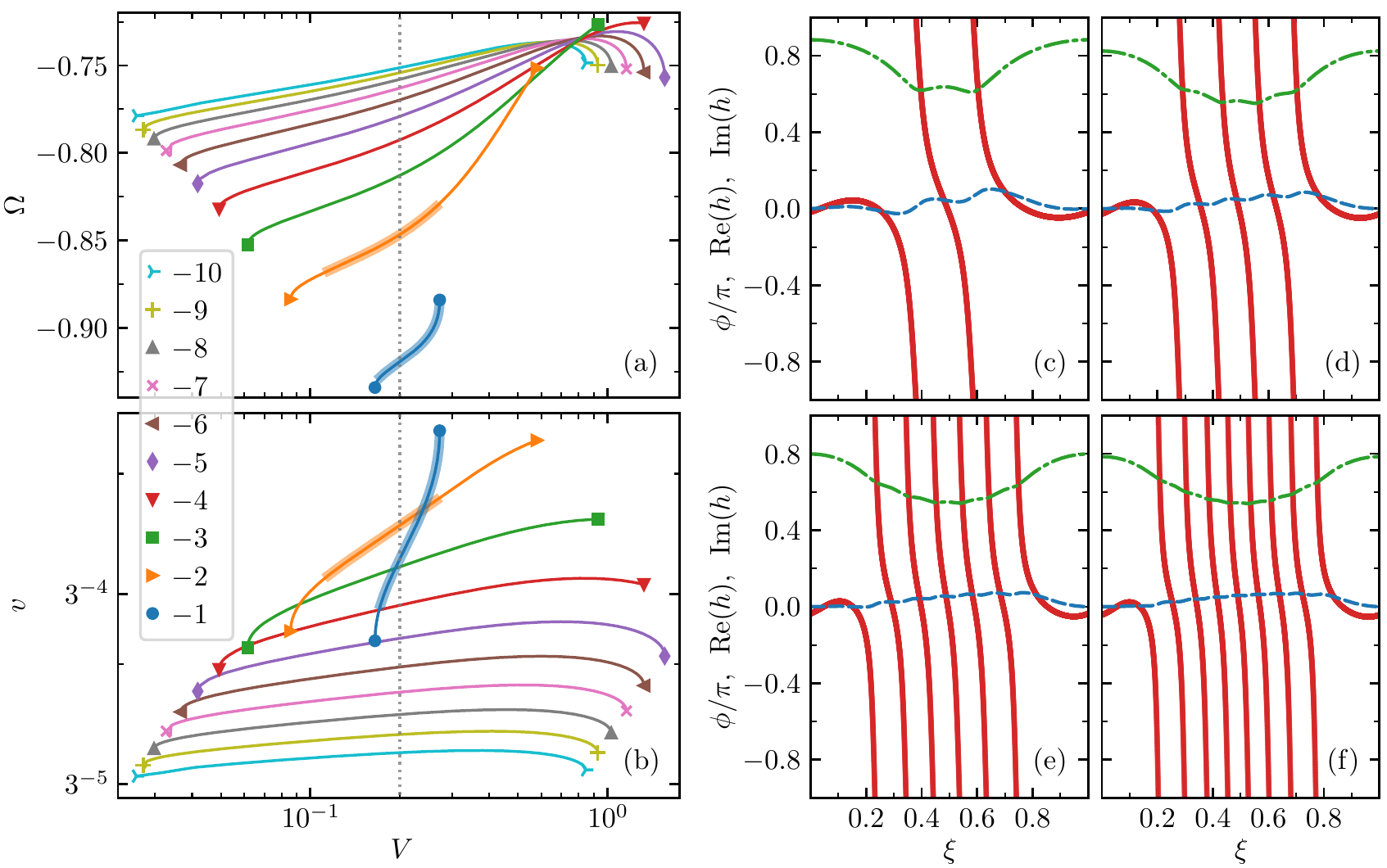} \vspace{-2.5mm}
\caption{The same as in Fig.~\ref{fig:branches-1} but for $\kappa=5$ and $\alpha=1.5$ (such a set of values corresponds to the parameters of Sec.~\ref{sec:il}).
In this case, the traveling patterns with $M=-1$ and $M=-2$ are stable.
Examples of the traveling wave patterns depicted on panels (c){\,}--{\,}(f) correspond to the advection parameter $V=0.2$ (dotted vertical lines on panels (a) and (b)) and the following values of $M$: (c) $M=-2$, (d) $M=-4$, (e) $M=-6$, and (f) $M=-8$. \vspace{-2.5mm}}
\label{fig:branches-2}
\end{figure}

One of the fixed points of the Poincar\'{e} map is stable and the other is unstable.
We take the stable fixed point as an initial condition to reconstruct the phase distribution $\phi(\xi)$ corresponding to the periodic solution of Eq.~\eqref{eq:eqs-qph-qp} for a given approximation for the structure of acting mean field $h(\xi)$ and current values $\Omega$ and $v$ of the traveling pattern parameters.

Next, we use the profile $\phi(\xi)$ obtained in the above way to calculate a new approximation for the complex function $h(\xi)$.
In order to do this, we evaluate the convolution integral 
\begin{equation} \label{eq:convolution}
h(\xi)=\int_{0}^{1}\!G(\xi-\tilde{\xi})e^{i\phi(\tilde{\xi})}d\tilde{\xi}
\end{equation}
employing the fast Fourier transform method.
Here, the kernel $G(\xi)$ is the asymmetric in space exponential-type kernel~\eqref{eq:egf}.
Then, we find the point $\xi_{0}$ where the derivative $\phi'(\xi)$ of the phase profile vanishes, i.{\,}e. $\phi'(\xi_{0})=0$.
According to Eq.~\eqref{eq:eqs-qph-qp}, this allows one to compute a new value of $\Omega$ approximating the genuine rotation frequency of the traveling wave pattern.
Integrating Eq.~\eqref{eq:eqs-qph-qp} over the interval ${0}\leq\xi\leq{1}$ and taking into account the boundary conditions, we obtain a formula for the velocity $v$ which can be used to refine its current value for the next step of the auxiliary iterative procedure.
As a result, we arrive at the following update rules for $\Omega$ and $v$:
\begin{equation} \label{eq:omega-v}
\Omega=\mathrm{Im}\Bigl(h(\xi_{0}) e^{-i\phi(\xi_{0})-i\alpha}\Bigr), \quad
v=\frac{1}{2\pi{M}}\left[\Omega-\int_{0}^{1}\!\mathrm{Im}\Bigl(h(\tilde{\xi}) e^{-i\phi(\tilde{\xi})-i\alpha}\Bigr)d\tilde{\xi}\right].
\end{equation}
After that, in order to find a good initial approximation close to the genuine values of $\mathcal{Q}$, $\mathcal{R}$, $\mathcal{S}$, $\Omega$, and $v$ which is  appropriate for the shooting approach, we repeat several times the above iteration scheme to get quantitatively acceptable profiles $\phi(\xi)$ and $h(\xi)$.

\begin{figure}[t]
\centering
\includegraphics[width=\textwidth]{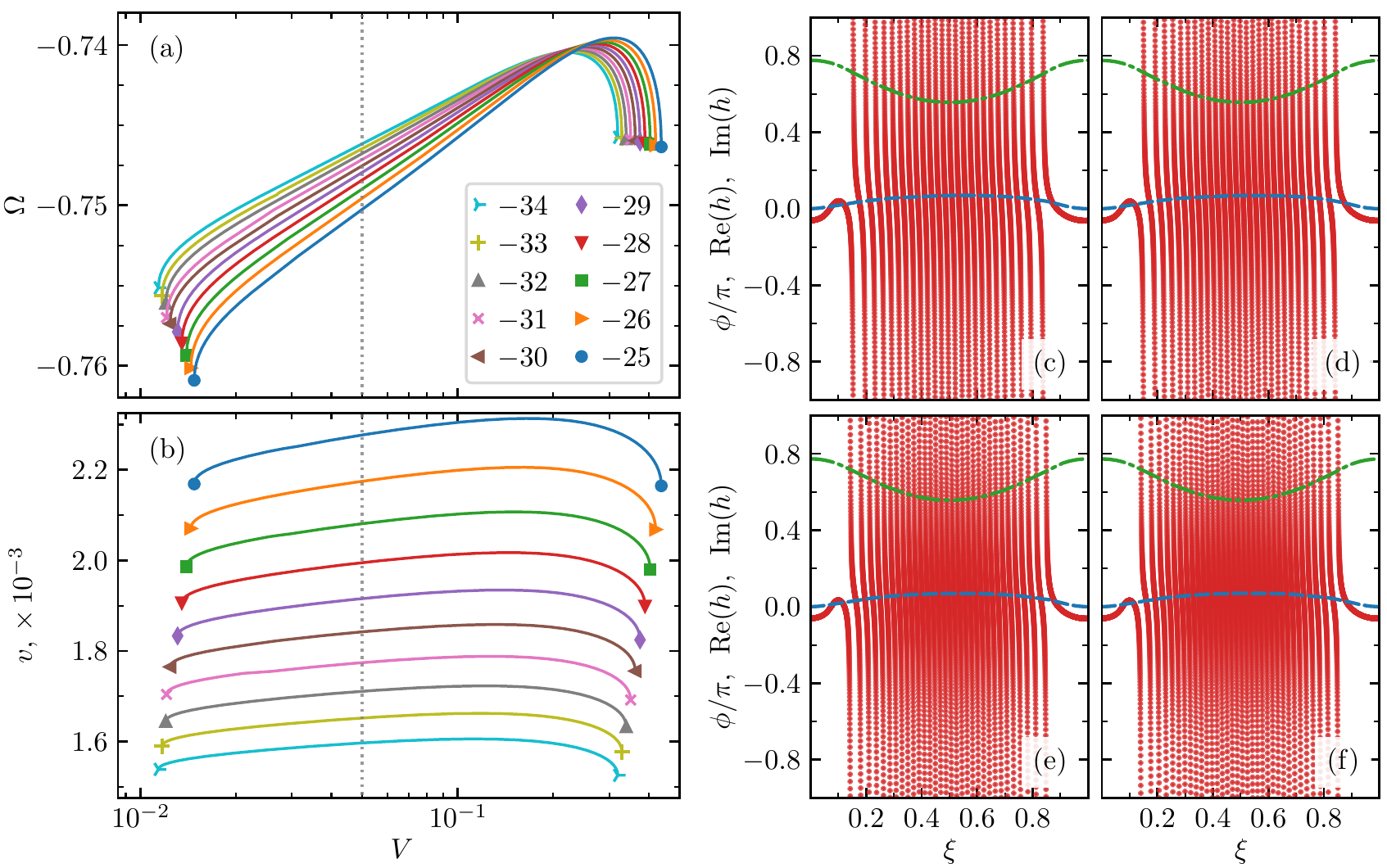} \vspace{-2.5mm}
\caption{The same as in Fig.~\ref{fig:branches-2} but for larger absolute values of the phase shift number $M$.
Examples of the traveling wave profiles shown on panels (c){\,}--{\,}(f) correspond to the advection parameter $V=0.05$ (dotted vertical lines on panels (a) and (b)) and the following values of $M$: (c) $M=-25$, (d) $M=-27$, (e) $M=-32$, and (f) $M=-34$. \vspace{-2.5mm}}
\label{fig:branches-3}
\end{figure}

\vspace{-1.25mm}
\subsection{Results: traveling wave profiles}
In Figures \ref{fig:branches-1}{\,}-{\,}\ref{fig:branches-3} we report branches of the solutions for different
values of the parameters $V,\kappa,\alpha,M$, obtained by virtue of the combination of the iterative procedure above with the Newton-Raphson method for continuation
along parameter $V$.  For each set $\kappa,\alpha,M$, the branch of TW solutions is 
limited  in the advection parameter $V$. These ranges of $V$ shift to smaller values for larger $|M|$: the traveling waves 
with larger number of phase rotations  exist for smaller advection terms, and have smaller velocities. It appears 
that traveling waves can be found for very small values of $V$, although this would require considering profiles with 
a very large number of phase shifts $|M|$ (e.{\,}g., see Fig.~\ref{fig:branches-3}).  

\begin{figure}[t]
\centering
\includegraphics[width=0.99\textwidth]{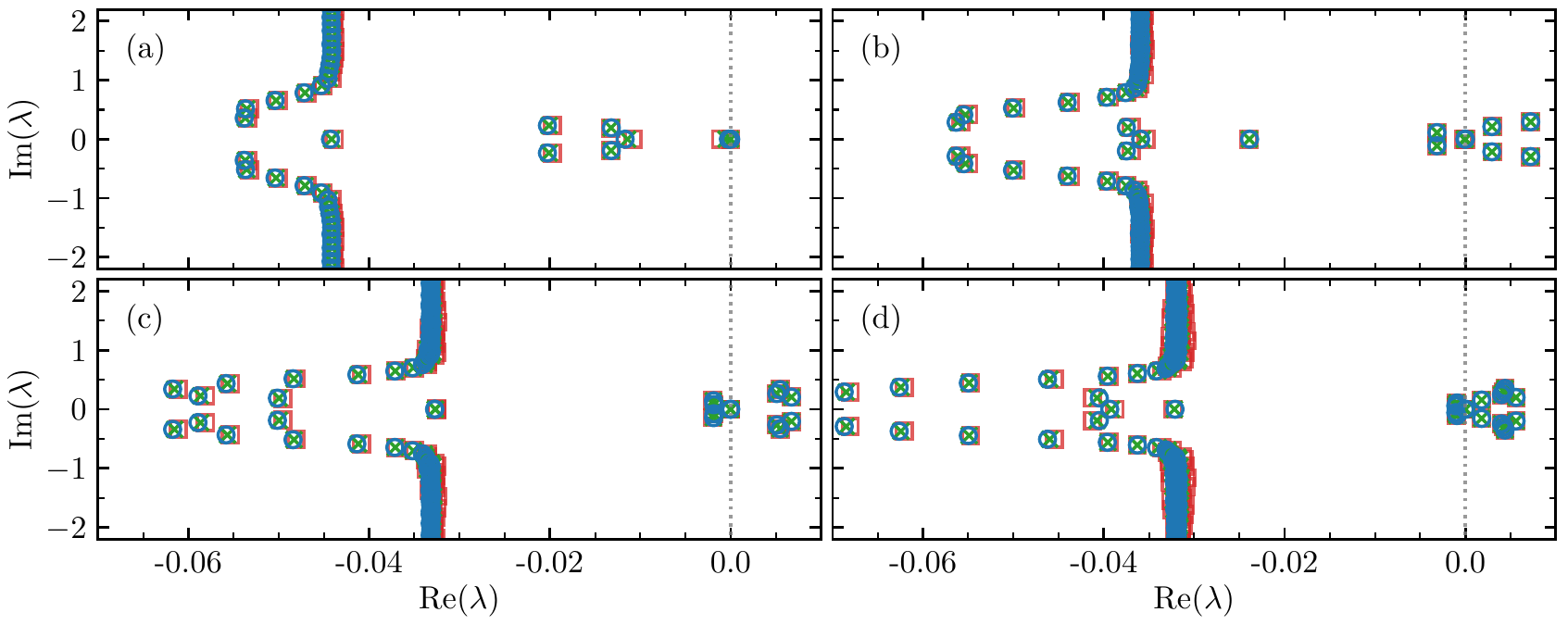} \vspace{-2.5mm}
\caption{Linear stability spectrum $\lambda$ of traveling wave solutions obtained for the corresponding set of parameters $\alpha=1.5$, $\kappa=5$, $V=0.5$, and the following values of phase shift number $M$: (a) $M=-2$ (here the solution is stable), (b) $M=-4$ (here the solution is unstable with four unstable complex eigenvalues), (c) $M=-6$ (here the solution is unstable with six unstable complex eigenvalues), and $M=-8$ (here the solution is unstable with eight unstable complex eigenvalues). The profiles of phase distributions $\phi(\xi)$ and the attributed acting mean field structures $h(\xi)$ are shown in Fig.~\ref{fig:branches-2}{\,}(c){\,}--{\,}(f). Different symbols and colors mean different discretizations: red square markers represent results of calculations employing $512$ discrete points, green crosses correspond to $1024$ discrete points, and blue circles represent results of calculations for $2048$ discrete points. \vspace{-2.5mm}}
\label{fig:stab}
\end{figure}

\vspace{-1.25mm}
\subsection{Results: stability of traveling waves}
In Figs.~\ref{fig:branches-1}{\,}-{\,}\ref{fig:branches-3}  we also depict the stability of the solution. It has been evaluated as follows.
First, we  re-write the equation in the traveling reference frame, where the traveling wave is a stationary solution:
 \begin{equation}
\frac{\partial\varphi(\xi,t)}{\partial{t}}=v \frac{\partial\varphi(\xi,t)}{\partial\xi}-\Omega+\int_{0}^{1}\!G(\xi-\tilde{\xi})\sin\!\left(\varphi(\xi,t)-\varphi(\tilde{\xi},t)-\alpha\right)d\tilde{\xi}.
\end{equation}
Linearization $\varphi(\xi,t)=\phi(\xi,t)+\hat{\varphi}(\xi,t)$ yields
\begin{equation}
\frac{\partial\hat{\varphi}(\xi,t)}{\partial{t}}=v \frac{\partial\hat{\varphi}(\xi,t)}{\partial\xi}-\Omega+\int_{0}^{1}\!G(\xi-\tilde{\xi})\cos\!\left(\phi(\xi)-\phi(\tilde{\xi})-\alpha\right)\left(\hat{\varphi}(\xi,t)-\hat{\varphi}(\tilde{\xi},t)\right)d\tilde{\xi},
\end{equation}
where $\hat{\varphi}(\xi,t)$ describes $\xi$-periodic small deviations from the traveling wave profile $\phi(\xi)$.
Now, with a spatial discretization we get a matrix, eigenvalues $\lambda$ of which can be found numerically. 
Relevant eigenvalues have finite imaginary parts (there are also some spurious eigenvalues with very large imaginary parts that are not relevant).
Reliability of this approach is confirmed by the overlap of found eigenvalues for different discretizations, as shown at Fig.~\ref{fig:stab}.

In the explored range of the parameters we have found stable traveling waves with $|M|=1,2$ only, all waves with larger phase shifts are unstable.
This observation corresponds to the statistical analysis of different asymptotic states in Fig.~\ref{fig:lt}: in the range of parameters where the traveling wave with $|M|=1$ is stable, it almost always appears after a long evolution of a transient turbulent chimera.
For other parameter values, either the turbulent chimera persists, or a synchronous state establishes.

\section{Discussion} \label{sec:dis}
Summarising, we have studied a one-dimensional medium of identical oscillators, coupled via an external field which is subject to diffusion and advection.
This setup generalizes the KB model, where only diffusion is present.
Due to advection, the coupling is asymmetric, and a chimera pattern starts to move.
We have demonstrated, that for a dense system (large number of oscillators), one observes strong correlations between the oscillators not only in the former synchronous region (where the phases are nearly equal to each other), but also in the former disordered domain, which in the moving case looks locally like a continuous profile of phases with a nearly constant gradient.
Such profiles indeed, as shown in section~\ref{sec:tw}, can be found as traveling waves in the system.
However, these regular waves are typically unstable, and weakly turbulent regimes where together with continuous profiles phase slips and small disordered regions exist, are observed.

Appearance of regularity of the phase profile due to motion is not surprising, if one considers stability properties of the oscillators dynamics.
In a standing chimera, the oscillators in the synchronous domain have stable dynamics, while those in the disordered domain are marginally stable (their Lyapunov exponent vanishes in the thermodynamic limit).
When synchronous and asynchronous patches start to move, each oscillator experiences epochs of stability and marginality, so that in average one has stability for all oscillators.
This stability means that neighbouring elements are close to each other, because they are driven by close forces.
The more dense are the elements, the more close are the forces acting on the nearest neighbours, and the more visible is the coherence between them.
We have characterized this local coherence with the spatial correlation function, which for the moving chimera attends values close to one for the nearest neighbours, while for a standing chimera the correlations do not exceed $1/3$.

We have found exact traveling wave solutions and studied their stability.
Only in some ranges of parameter stable waves have been found, and for these parameters such waves typically appear as attractors after a long chaotic chimera transient.
In other parameter domains no stable traveling waves exist, and although such a wave can be observed during initial evolution starting from the standing chimera, due to instability it is destroyed and a weakly turbulent chimera establishes.
We have checked that this turbulence is not a finite-size effect (like chaos in a standing chimera), by showing that the largest Lyapunov exponent of the system remains size-independent starting from a certain number of units $N$ (Fig.~\ref{fig:lyap}).
The weak turbulent chimera exists for  long time intervals, but in some regions of parameters we observed a transition to a fully synchronous state (or to a regular twisted wave). It is, however, not completely clear if for these values of parameters there is a bistability of chimera and synchrony, or in all cases the chimera is a transient, although with a very long lifetime.

Finally, we would like to mention that similar features can be found in the chimera setup suggest by Abrams and Strogatz~\cite{Abrams-Strogatz-04}; these results will be reported elsewhere.
\acknowledgments 
This paper was supported by the Russian Science Foundation (Secs.~\ref{sec:il} and~\ref{sec:tw}, Grant No.~22-12-00348), and the Scientific and Education Mathematical Center ``Mathematics for Future Technologies'' (Sec.~\ref{sec:bm}, Project No. 075-02-2022-883).
We thank O. Omelchenko, E. Knobloch, and M. Bolotov for fruitful discussions.

\bibliography{chbibl}

\begin{thebibliography}{10}
\expandafter\ifx\csname urlstyle\endcsname\relax
  \providecommand{\doi}[1]{(doi:\discretionary{}{}{}#1)}\else
  \providecommand{\doi}{(doi:\discretionary{}{}{}\begingroup
  \urlstyle{rm}\Url)}\fi

\bibitem{Panaggio-Abrams-15}
Panaggio MJ, Abrams DM. 2015 Chimera states: coexistence of coherence and
  incoherence in networks of coupled oscillators.
\newblock \emph{Nonlinearity} \textbf{28}, R67--R87.

\bibitem{Omelchenko-18}
Ome{l'}chenko OE. 2018 The mathematics behind chimera states.
\newblock \emph{Nonlinearity} \textbf{31}, R121--R164.

\bibitem{Omelchenko-Knobloch-19}
Omel'chenko OE, Knobloch E. 2019 Chimerapedia:
  coherence{\textendash}incoherence patterns in one, two and three dimensions.
\newblock \emph{New Journal of Physics} \textbf{21}, 093034.

\bibitem{Kuramoto-Battogtokh-02}
Kuramoto Y, Battogtokh D. 2002 Coexistence of coherence and incoherence in
  nonlocally coupled phase oscillators.
\newblock \emph{Nonlinear Phenom. Complex Syst.} \textbf{5}, 380--385.

\bibitem{Tinsley_etal-12}
Tinsley MR, Nkomo S, Showalter K. 2012 Chimera and phase-cluster states in
  populations of coupled chemical oscillators.
\newblock \emph{Nature Physics} \textbf{8}, 662--665.

\bibitem{wickramasinghe2013spatially}
Wickramasinghe M, Kiss IZ. 2013 Spatially organized dynamical states in
  chemical oscillator networks: Synchronization, dynamical differentiation, and
  chimera patterns.
\newblock \emph{PloS one} \textbf{8}, e80586.

\bibitem{Martens_etal-13}
Martens EA, Thutupalli S, Fourri{\`e}re A, Hallatschek O. 2013 Chimera states
  in mechanical oscillator networks.
\newblock \emph{Proc. Natl. Acad. Sci.} \textbf{110}, 10563--10567.

\bibitem{Totz_etal-18}
Totz JF, Rode J, Tinsley MR, Showalter K, Engel H. 2018 Spiral wave chimera
  states in large populations of coupled chemical oscillators.
\newblock \emph{Nature Physics} \textbf{14}, 282.

\bibitem{Abrams-Strogatz-04}
Abrams DM, Strogatz SH. 2004 Chimera states for coupled oscillators.
\newblock \emph{Phys. Rev. Lett.} \textbf{93}, 174102.

\bibitem{Omelchenko_etal-08}
Omel'chenko OE, Maistrenko YL, Tass PA. 2008 Chimera states: The natural link
  between coherence and incoherence.
\newblock \emph{Phys. Rev. Lett.} \textbf{100}, 044105.

\bibitem{Laing-09}
Laing CR. 2009 The dynamics of chimera states in heterogeneous {K}uramoto
  networks.
\newblock \emph{Physica D} \textbf{238}, 1569 -- 1588.

\bibitem{Bordyugov-Pikovsky-Rosenblum-10}
Bordyugov G, Pikovsky A, Rosenblum M. 2010 Self-emerging and turbulent chimeras
  in oscillator chains.
\newblock \emph{Phys. Rev. E} \textbf{82}, 035205.

\bibitem{Omelchenko-13}
Omel'chenko OE. 2013 Coherence-incoherence patterns in a ring of non-locally
  coupled phase oscillators.
\newblock \emph{Nonlinearity} \textbf{26}, 2469.

\bibitem{xie2015chimera}
Xie J, Kao HC, Knobloch E. 2015 Chimera states in systems of nonlocal
  nonidentical phase-coupled oscillators.
\newblock \emph{Physical Review E} \textbf{91}, 032918.

\bibitem{Kemeth_etal-16}
Kemeth FP, Haugland SW, Schmidt L, Kevrekidis IG, Krischer K. 2016 A
  classification scheme for chimera states.
\newblock \emph{Chaos} \textbf{26}, 094815.

\bibitem{Smirnov-Osipov-Pikovsky-17}
Smirnov L, Osipov G, Pikovsky A. 2017 Chimera patterns in the
  {K}uramoto-{B}attogtokh model.
\newblock \emph{Journal of Physics A: Mathematical and Theoretical}
  \textbf{50}, 08LT01.

\bibitem{laing2015chimeras}
Laing CR. 2015 Chimeras in networks with purely local coupling.
\newblock \emph{Physical Review E} \textbf{92}, 050904.

\bibitem{Smirnov_etal-22}
Smirnov LA, Bolotov MI, Bolotov DI, Osipov GV, Pikovsky A. 2022
  Finite-density-induced motility and turbulence of chimera solitons.
\newblock \emph{New Journal of Physics} \textbf{24}, 043042.

\bibitem{Omelchenko_etal-10}
Omel'chenko OE, Wolfrum M, Maistrenko YL. 2010 Chimera states as chaotic
  spatiotemporal patterns.
\newblock \emph{Phys. Rev. E} \textbf{81}, 065201.

\bibitem{Xie_etal-14}
Xie J, Knobloch E, Kao HC. 2014 Multicluster and traveling chimera states in
  nonlocal phase-coupled oscillators.
\newblock \emph{Phys. Rev. E} \textbf{90}, 022919.

\bibitem{Omelchenko-19}
Omel’chenko O. 2019 Travelling chimera states in systems of phase oscillators
  with asymmetric nonlocal coupling.
\newblock \emph{Nonlinearity} \textbf{33}, 611.

\bibitem{Smirnov_Osipov_Pikovsky-18}
Smirnov LA, Osipov GV, Pikovsky A. 2018 Solitary synchronization waves in
  distributed oscillator populations.
\newblock \emph{Phys. Rev. E} \textbf{98}, 062222.

\bibitem{dudkowski2019traveling}
Dudkowski D, Czo{\l}czy{\'n}ski K, Kapitaniak T. 2019 Traveling chimera states
  for coupled pendula.
\newblock \emph{Nonlinear Dynamics} \textbf{95}, 1859--1866.

\bibitem{Kuramoto_etal-00}
Kuramoto Y, Nakao H, Battogtokh D. 2000 Multi-scaled turbulence in large
  populations of oscillators in a diffusive medium.
\newblock \emph{Physica A: Statistical Mechanics and its Applications}
  \textbf{288}, 244--264.

\bibitem{Tanaka-Kuramoto-03}
Tanaka D, Kuramoto Y. 2003 Complex {G}inzburg-{L}andau equation with nonlocal
  coupling.
\newblock \emph{Phys. Rev. E} \textbf{68}, 026219.

\bibitem{Shima-Kuramoto-04}
Shima SI, Kuramoto Y. 2004 Rotating spiral waves with phase-randomized core in
  nonlocally coupled oscillators.
\newblock \emph{Phys. Rev. E} \textbf{69}, 036213.

\bibitem{Bolotov_etal-22}
Bolotov DI, , Bolotov MI, Smirnov LA, Osipov GV, Pikovsky A. 2022
  Synchronization regimes in an ensemble of phase oscillators coupled through a
  diffusion field.
\newblock \emph{Radiophysics and Quantum Electronics} \textbf{64}, 709--725.

\bibitem{Wolfrum_Omelchenko-11}
Wolfrum M, Omel'chenko OE. 2011 Chimera states are chaotic transients.
\newblock \emph{Phys. Rev. E} \textbf{84}, 015201.

\bibitem{Omel-BrChimeras-2022}
Ome{l'}chenko OE. 2022 Mathematical framework for breathing chimera states.
\newblock \emph{J Nonlinear Sci} \textbf{32}, 22.

\bibitem{Marvel-2009}
Marvel SA, Mirollo RE, Strogatz SH. 2009 Identical phase oscillators with
  global sinusoidal coupling evolve by möbius group action.
\newblock \emph{Chaos: An Interdisciplinary Journal of Nonlinear Science}
  \textbf{19}, 043104.

\bibitem{Wilczynski-2008}
Wilczyński P. 2008 Planar nonautonomous polynomial equations: The riccati
  equation.
\newblock \emph{Journal of Differential Equations} \textbf{244}, 1304--1328.

\bibitem{Campos-1997}
Campos J. 1997 M\:{o}bius transformations and periodic solutions of complex
  riccati equations.
\newblock \emph{Bull. London Math. Soc.} \textbf{29}, 205.

\bibitem{Gong-2020}
Gong CC, Toenjes R, Pikovsky A. 2020 Coupled m\"{o}bius maps as a tool to model
  kuramoto phase synchronization.
\newblock \emph{Physical Review E} \textbf{102}, 022206.

\end{thebibliography}

\end{document}